\documentclass[prc,twocolumn,floatfix,nofootinbib,reprint]{revtex4-1}

\usepackage{graphicx}
\usepackage{amssymb,amsmath,amstext,amsthm,amsfonts}
\usepackage{slashed}

\newcommand{\onepion}{single-pion\ }

\newcommand{\dd}{\mathrm{d}}

\begin{document}

\title{Pion Production in high-energy neutrino reactions with nuclei}

\author{U. Mosel}
\affiliation{Institut fuer Theoretische Physik, Universitaet Giessen, D35392- Giessen, Germany}
\email[Contact e-mail: ]{mosel@physik.uni-giessen.de}

\begin{abstract}
\begin{description}
\item[Background] A quantitative understanding of neutrino interactions with nuclei is needed for precision era neutrino long baseline experiments (MINOS, NOvA, DUNE) which all use nuclear targets. Pion production is the dominant reaction channel at the energies of these experiments.

\item[Purpose] Investigate the influence of nuclear effects on neutrino-induced pion production cross sections and compare predictions for pion-production with available data.

\item[Method] The Giessen Boltzmann--Uehling--Uhlenbeck (GiBUU) model is used for the description of all incoherent channels in neutrino-nucleus reactions.

\item[Results] Differential cross sections for charged and neutral pion production for the MINER$\nu$A neutrino and antineutrino flux are calculated. An estimate for the coherent cross section is obtained from a comparison of data with theoretical results for incoherent cross sections. The invariant mass ($W$) distribution of the $\Delta$ resonances produced is analyzed.

\item[Conclusions]  Final state interactions affect the pion kinetic energy spectra significantly. The data for charged pion production at MINER$\nu$A are compatible with the results of calculations using elementary data taken from an old Argonne National Laboratory experiment. Remaining differences for charged pion production can be attributed to coherent production; the data for antineutrino induced neutral pion production, where no coherent contribution is present, are reproduced quite well. The analysis of $W$-distributions shows that sharp cuts on experimentally reconstructed invariant masses lead to shape-distortions of the true $W$ distributions for nuclear targets.
\end{description}
\end{abstract}

\date{\today}

\maketitle

\section{Introduction}
The planned deep underground neutrino experiment (DUNE) will operate in a neutrino beam the properties of which are similar to the presently used NUMI beam at Fermilab. Already ongoing and planned experiments in that beam offer the possibility for precision measurements of the differential cross sections for quasielastic scattering (QE) and single- and multi-pion production, as a function of (reconstructed) neutrino energy. These cross sections are essential for an understanding of the interactions of the incoming neutrinos with the nuclear targets (C, O, Ar) used in these experiments. A quantitative understanding of these cross sections is essential for the energy reconstruction and - consequently - for a precise extraction of oscillation parameters from observed event rates \cite{Mosel:2013fxa}. This extraction has to rely on generators and thus the cross sections measured now in experiments such as, e.g., MINER$\nu$A could be a basis for a fine-tuning of event generators \cite{Minerva:2005} for use at the DUNE.

All events following from a  neutrino-nucleus interaction can be grouped into two general classes. The first one is that of quasielastic scattering (QE), including many-body interactions. The second class is connected with pion production, either through resonances, $t$-channel background processes or deep inelastic scattering (DIS). At the presently running MINER$\nu$A experiment the pion production channels make up for about 2/3 of the total inclusive cross section while the QE events account only for about 1/3 \cite{Mosel:2015oda}. Both event types are entangled with each other and cannot be separated by purely experimental means. Separation is only possible with the help of an event generator and, therefore, is necessarily model-dependent. Recent calculations of the inclusive QE response of nuclei \cite{Benhar:2010nx,Amaro:2011qb,Martini:2011wp,Nieves:2011yp,Lovato:2014eva} in their comparison with experiment thus have to rely on the accuracy with which pion production is described by the generator used in extracting the data.

It is, therefore, essential to obtain a quantitative understanding of the dominant reaction channels at MINER$\nu$A and DUNE; these are those connected with pion production. It is the purpose of the present paper to present results of calculations for pion production at the MINERV$\nu$A experiment. While first results were already contained in \cite{Lalakulich:2012gm,Mosel:2014lja} in the present paper now a detailed comparison with MINER$\nu$A data is performed using their flux and invariant mass cuts. Over the last few years valuable insight into the neutrino-induced pion production on nuclei at lower energies has been obtained mainly from the MiniBooNE experiment \cite{AguilarArevalo:2010bm,AguilarArevalo:2010xt,Lalakulich:2012cj,Hernandez:2013jka}. Therefore, also the consistency of its data with those obtained at the higher energies of the MINER$\nu$A experiment is discussed in the present paper. Finally, also some problems connected with kinematical cuts in the experiment are investigated.

\section{Method}

The calculations are performed within the transport theoretical framework GiBUU \cite{gibuu}. GiBUU approximately factorizes the reaction into a very first interaction of the incoming neutrino with a bound and Fermi-moving nucleon and the following final state interactions (FSI)\footnote{I denote by 'final state interactions' only the secondary and following collisions. This is different from the nomenclature often used in studies of inclusive cross sections for QE where FSI denote the potentials felt by the outgoing nucleon in the final state of the initial reaction.}. The latter are described by a numerical implementation of the Kadanoff-Baym equations in the gradient approximation \cite{Kad-Baym:1962}, using the Botermans-Malfliet approximation for off-shell transport \cite{Botermans:1990qi}. More details about this treatment of final state interactions can be found in Ref.\ \cite{Buss:2011mx}.

The groundstate of the target nucleus is assumed to be that of a \emph{local} relativistic Fermi gas with the nucleons being bound in a coordinate- and momentum-dependent potential that has been fitted to equation of state and effective mass data \cite{Welke:1988zz,Gale:1989dm}. The hole spectral function is given by
\begin{widetext}
\begin{equation}
P_h(\mathbf{p},E) = g \int\limits_{\rm nucleus} \!\!\!\dd^3r \,\Theta\left[p_\mathrm{F}(\mathbf{r}) - |\mathbf{p}|\right] \Theta(E) \delta\left(E - m^*(\mathbf{r},\mathbf{p}) + \sqrt{\mathbf{p}^2 + {m^*}^2(\mathbf{r},\mathbf{p})}\right)~;
\end{equation}
\end{widetext}
here $p_\mathrm{F}(\mathbf{r})$ is the local Fermi momentum  given by the local Thomas Fermi model and $g$ is a degeneracy factor. In this spectral function all effects of the nucleon potential are assumed to be contained in the effective mass $m^*$ \cite{Buss:2011mx} which depends on location and momentum of the nucleon. The corresponding momentum distribution approximates that obtained in state-of-the-art nuclear many-body theory calculations quite well; see Fig. 4 in \cite{Alvarez-Ruso:2014bla}. The initial interaction rates are calculated for a nucleon at rest; the results are then boosted to the local rest frame of the Fermi-moving target nucleon.

The \onepion production cross section at fixed neutrino energy is then given by (cf.\ \cite{Leitner:2006ww} where all the details can be found)
\begin{widetext}
\begin{equation}   \label{piprod}
\dd \sigma^{\nu A \to \ell'X\pi} =  g \int\limits_{\rm nucleus} \!\!\!\dd^3r \,\int \frac{\dd^3p}{(2\pi)^3} \Theta\left[p_\mathrm{F}(\mathbf{r}) - |\mathbf{p}|\right] f_{\rm corr}\, \dd\sigma^{\rm med} \, \mathcal{P}_{\rm PB} (\mathbf{r},\mathbf{p}) F_\pi(\mathbf{q}_\pi, \mathbf{r}) ~.
\end{equation}
\end{widetext}
Here $f_{\rm corr}$ is a flux correction factor, $f_{\rm corr} = (k \cdot p)/(k^0p^0)$; $k$ and $p$ denote the four-momenta of the neutrino and nucleon momentum, respectively. $\mathcal{P}_{\rm PB}(\mathbf{r},\mathbf{p})$ describes the Pauli-blocking and the factor $F_\pi(\mathbf{q}_\pi, \mathbf{r})$ in (\ref{piprod}) describes the effects of all the FSI contained in GiBUU. For pions the latter involve elastic and inelastic scattering as well as pion absorption through two-body and three-body processes of the types $N^* + N \to N + N$ and $N^* + N + N \to N + N + N$; here $N^*$ stands for the $\Delta$ resonance and higher excitations of the nucleon. In GiBUU these resonances are treated explicitly; they become excited and then are being propagated until they decay or collide with other nucleons; this is described in some detail in \cite{Buss:2011mx}. The time-development of the $\pi-N-\Delta$ dynamics in the nucleus is determined by the resonance widths and collision rates alone. In the resonance region there is no room for a further free parameter, such as a formation time used in other generators \cite{Golan:2012wx}. The use of formation times, during which interactions of the produced particle are prohibited, introduces a new parameter and a corresponding arbitrariness into the description of pion production. Only in the DIS part, above invariant nucleon masses of about 2 GeV, the concept of a formation time makes sense since it accounts for the widths of high lying, no longer separable excitations of the nucleon. Even then, cross sections of the produced particles should rise with time until the final hadron has fully been formed \cite{Gallmeister:2007an}.

The cross section $\dd\sigma^{\rm med}$ in Eq.\ (\ref{piprod}) stands for the pion production inside the nuclear medium. Pions can be produced either through nucleon resonances and $t$-channel (background) processes or through DIS. The latter, denoted in the following by $\dd\sigma_{\rm DIS}$, is obtained from the string-fragmentation model PYTHIA \cite{Sjostrand:2006za}. GiBUU smoothly switches over been these two pictures around an invariant mass of the nucleon of 2 GeV. Below this energy all nucleon resonances are treated with their correct pion decay branches taken from the PDG \cite{Amsler:2008zzb}. In this mass region the pion production cross section consists of a resonance and of a background contribution which have to be added coherently in order to obtain the full cross section. For the background contribution a form taken from an effective field theory treatment of pion production up the $\Delta$ resonance region \cite{Hernandez:2007qq,Lalakulich:2010ss} is used. No background contributions for the higher resonances are taken into account. Since transport (or any Monte Carlo code) cannot handle the coherence of resonance and background amplitudes the coherent sum is split up into a resonance cross section and one for the background which contains both the squared background amplitude and the interference term. Background pions are then produced locally, without any time-delay.  Because GiBUU describes only incoherent processes it does not contain a coherent pion-production cross section; the latter would require a quantum mechanical description.

The resonance production cross section is given by
\begin{equation}       \label{resprod}
\frac{\dd\sigma^{\rm med}}{\dd\omega \dd\Omega'} = \frac{|\mathbf{k}'|}{32 \pi^2} \frac{\mathcal{A}^{\rm med}(p')}{[(k \cdot p)^2 - m_\ell^2M^2]^{1/2}} \left| \mathcal{M}_R \right|^2 ~.
\end{equation}
Here $M$ is the nucleon mass, $p$ denotes the nucleon's four-momentum, $p'$ that of the outgoing resonance and $k$ and $k'$ that of the initial and final state lepton, respectively. The quantities $\omega$ and $\Omega'$ give the energy transfer and the scattering angle of the outgoing lepton, respectively. The in-medium spectral function of the resonance is denoted by $\mathcal{A}^{\rm med}(p')$.

The cross section for resonance formation (\ref{resprod}) contains the square of an invariant matrix element $\mathcal{M}_R$ that is obtained by contracting the lepton tensor with the hadron tensor. For the latter one has
\begin{equation}
H^{\mu \nu} = \frac{1}{2} {\rm Tr} \left[\slashed{p} + M) \Gamma^{\alpha \mu} \Lambda_{\alpha \beta} \Gamma^{\beta \nu}\right]
\end{equation}
where $\Lambda_{\alpha \beta}$ is (for the $\Delta$) the spin-3/2 projector and the vertex factor $\Gamma^{\alpha \mu}$ is given by
\begin{equation}
\Gamma^{\alpha \mu} = \left[ V^{\alpha \mu} - A^{\alpha \mu} \right] \gamma^5
\end{equation}
for a positive parity resonance. The vector part $V$ is taken from the MAID analysis of electron scattering data \cite{MAID} so that the data for electro excitation of nucleon resonances are reproduced by construction. For the axial part $A$ the spin-3/2 transition current contains in principle four independent axial form factors $C^A$, but the presently available data do not allow to determine them separately.  Therefore, one of them($C_3^A$) is is set equal to zero
\begin{widetext}
\begin{equation}
A^{\alpha \mu} = -  \left(\frac{C_4^A(Q^2)}{M^2}(g^{\alpha \mu} q\cdot p' - q^\alpha p'^\mu) + C^A_5(Q^2) g^{\alpha \mu} + \frac{C^A_6(Q^2)}{M^2} q^\alpha q^\mu \right) \gamma^5
\end{equation}
\end{widetext}
with the further simplification $C^A_4 = - C_5^A/4$ and $C^A_6 = C_5^A M^2/(Q^2 + m_\pi^2)$. More details can be found in Ref.\  \cite{Leitner:2006ww}.

The absolute strength of the resonance contributions determined by $C^A_5(0)$ is obtained by fitting the available pion production data on an elementary target. The two datasets available are those obtained at Argonne National Lab (ANL) \cite{Radecky:1981fn} and Brookhaven National Lab (BNL) \cite{Kitagaki:1986ct}. These two datasets differ in the relevant energy regime, with the BNL dataset being higher than the ANL one by about 25\%. This introduces a corresponding uncertainty into the calculations for nuclear targets. Earlier it had been argued that this difference is due to flux-uncertainties \cite{Graczyk:2009qm}. The authors of \cite{Lalakulich:2010ss}, in a detailed study of the consistency of the various isospin channels and the measured $d\sigma/dQ^2$, concluded that the BNL data were too high. This has very recently been verified by the authors of \cite{Wilkinson:2014yfa}. In a reanalysis of the old data that fixes the flux with the help of the QE cross section it was shown that at least for the $\pi^+$ channel the ANL data  were preferable and also consistent with other data from CERN \cite{Wilkinson:2014yfa}. The reanalysis of Wilkinson et al.\ \cite{Wilkinson:2014yfa} seems to settle the question for the correct elementary cross section. However, most recently a new theoretical calculation of pion production on deuterium has shown that even in this small system FSI can play a significant role \cite{Wu:2014rga}. This then affects the extraction of cross sections for $p$ and $n$ targets from data obtained with a D target. There is thus still some uncertainty left on the elementary pion production cross section.

The extensive pion production data from the MiniBooNE experiment \cite{AguilarArevalo:2010bm,AguilarArevalo:2010xt} obtained on a CH$_2$ target are consistently higher than the ones calculated within GiBUU, both for the ANL and -- less so -- for the BNL input cross sections \cite{Lalakulich:2012cj,Hernandez:2013jka}. Motivated by the reanalysis by Wilkinson et al \cite{Wilkinson:2014yfa} now the ANL dataset is used as default input into GiBUU.  In order to illustrate the sensitivity of the results for a nuclear target to the remaining uncertainties in the nucleon cross section, however, also some results obtained with the BNL input are shown.

To obtain the pion production cross section the resonance formation cross section is multiplied with the the branching ratio for decay into the $\pi N$ channel
\begin{equation}
\frac{\dd\sigma^{\rm med}}{\dd\omega \dd\Omega' \dd\Omega_\pi^{CM}} =  \frac{1}{4 \pi} \frac{\dd\sigma^{\rm med}}{\dd\omega \dd\Omega'} \frac{\Gamma_{R \to N\pi}}{\Gamma_{\rm tot}}
\end{equation}
Here it is assumed that the decay of the resonance happens isotropically in the rest frame of the resonance. This is not a good approximation for an isolated $\Delta$ resonance. Therefore, for the FSI process $\pi + N \to \Delta \to \pi + N$ a more sophisticated parametrization of the $p$-wave behavior of the $\Delta$ decay is used as explained in detail in \cite{Buss:2006yk}. Although most of the primary reactions populate higher lying states via DIS or higher resonances, the $\Delta$ resonance plays quite an essential role in the observable pion spectrum as will be discussed later.

The in-medium effects for pion production are contained both in the spectral function $\mathcal{A}$ in Eq.\ (\ref{resprod}) as well as in the branching ratio for the resonance decay into $\pi N$ where the final nucleon state may be Pauli-blocked. The final state nucleon is bound in a momentum and coordinate dependent potential which -- through energy- and momentum-dependence -- affects the decay width. The spectral function contains a collisional broadening which is taken from intensive investigations of $\Delta$ properties inside the nuclear medium by the Valencia group \cite{Oset:1987re}. To obtain the correct spectral distribution of pions it is essential to maintain  consistency between this collisional broadening and the actual collision rates determined by the collision cross sections encoded in the generator.

The theory described so far has been extensively tested with the help of photon- \cite{Krusche:2004uw} and electron-induced \cite{Kaskulov:2008ej} pion production data. This is not the case for the recent work of
Yu et al.\ \cite{Yu:2014yja} which uses the Adler-Nussinov-Paschos model for a description of pion production in the MiniBooNE and MINER$\nu$A experiments. This model does not describe the $\pi-N-\Delta$ dynamics correctly and indeed the authors of \cite{Yu:2014yja} themselves ascribe the failure of the model to describe the MINER$\nu$A data to the absence of all pion-nucleon interactions, except for absorption, in that model.

For the comparison with the experimental neutrino data the fixed-energy cross sections described above are folded with the flux
\begin{equation}
\langle d\sigma \rangle = \int_{1.5 \rm GeV}^{10 \rm GeV} \rm{d}E_\nu \, \phi(E_\nu) \left(\dd\sigma^{\rm med}(E_\nu) + \dd\sigma_{\rm DIS}(E_\nu) \right)
\end{equation}
where $\phi(E_\nu)$ is the incoming energy distribution (the 'flux'), normalized to 1. The integration boundaries 1.5 and 10 GeV are those of the experimental analysis.  Furthermore, for the charged-pion cross sections a cut on the incoming invariant mass $W < 1.4$ GeV has been imposed, as in the experimental analysis. All results shown in this paper were obtained with the MINER$\nu$A flux for a CH target. All cross sections are given per nucleon, i.e.\ the total cross section for CH is divided by 13.

\begin{figure}[htb]
 \includegraphics[angle=-90,width=0.5\textwidth]{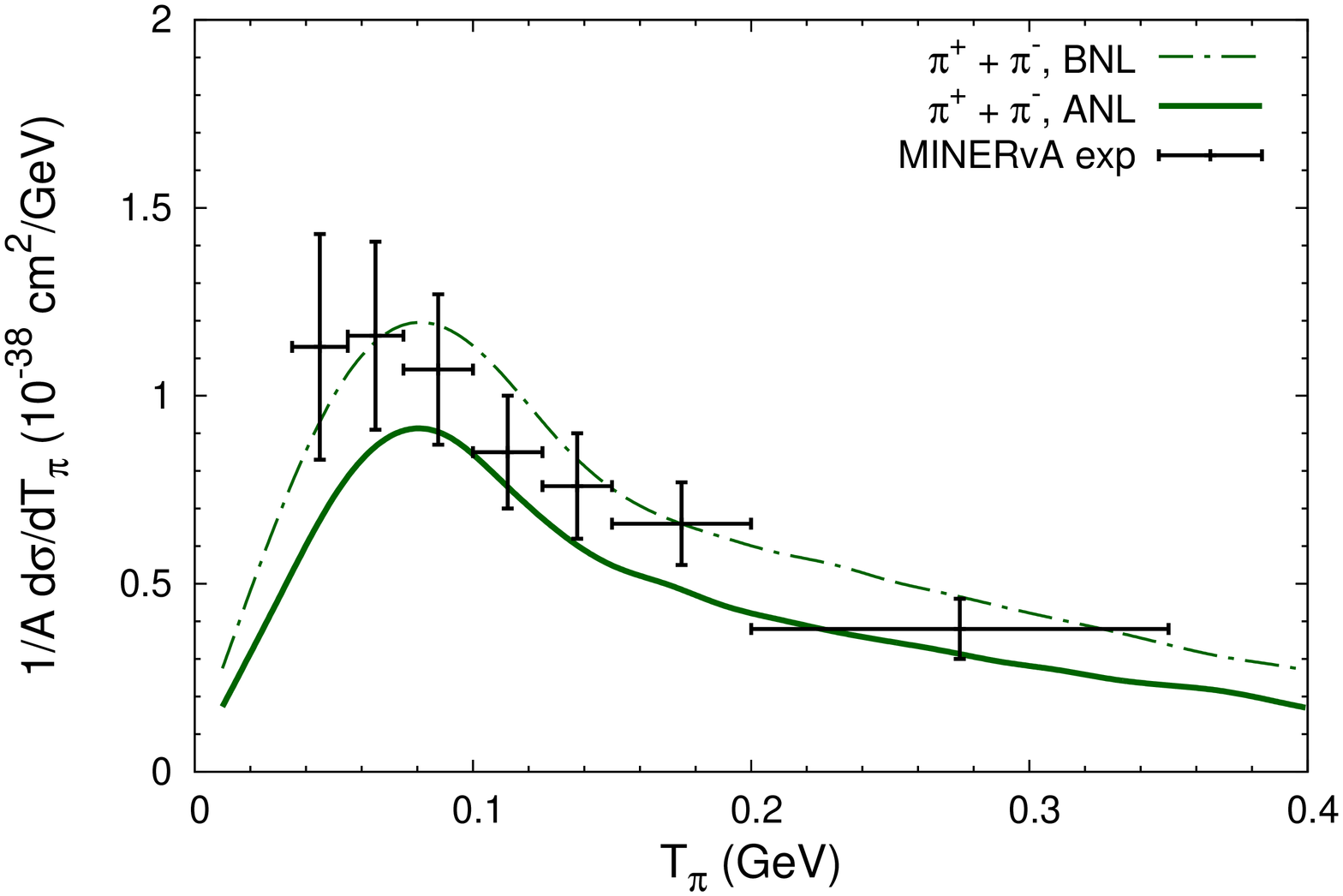}
 \caption{(color online) Kinetic energy distributions per nucleon of incoherently produced charged pions on a CH target in the MINER$\nu$A neutrino flux. The solid curve gives the results obtained with the ANL cross sections as elementary input. The dashed-dotted curve gives the same distribution calculated with the BNL elementary cross sections.  The data are from Ref.\ \cite{Eberly:2014mra}.} \label{fig:Tpi}
 \includegraphics[angle=-90,width=0.5\textwidth]{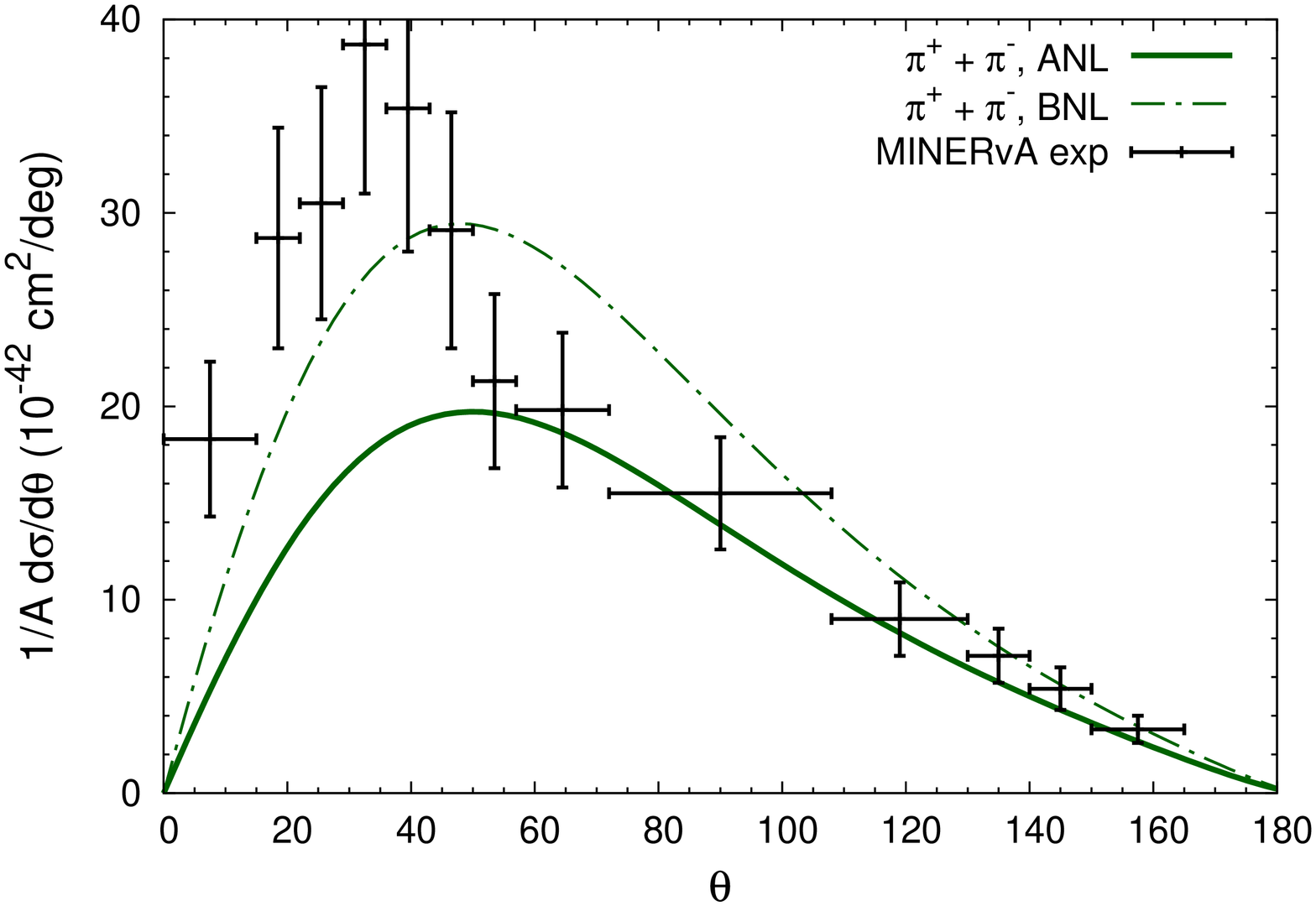}
 \caption{(color online) Angular distribution of incoherently produced charged pions in the MINER$\nu$A experiment (data from \cite{Eberly:2014mra}). The solid curve gives results of a calculation with the ANL input, the dash-dotted curve has been obtained with the BNL input.} \label{fig:piang}
\end{figure}

\section{Pion Production at MINER$\nu$A}

\subsection{Charged pion production in the neutrino beam}

In Fig.\ \ref{fig:Tpi} the calculated kinetic energy distribution for the sum of $\pi^+$ and $\pi^-$ is shown in comparison with the data \cite{Eberly:2014mra}. The solid curve, calculated with the ANL input, follows the shape reasonably well. The biggest disagreement shows up for kinetic energies $T_\pi < 0.1$ GeV; for the higher energies the calculation lies at the lower end of the error bars so that the disagreement there may not be significant.

Before discussing the difference between theoretical and experimental kinetic energy distributions further in Fig.\ \ref{fig:piang} the calculated angular distribution is shown in comparison with the data. It is evident now that the disagreement noticed for the kinetic energy distribution is localized at forward angles $\theta < 50^\circ$. At about $30^\circ$ the calculated cross section amounts to only about 1/2 of the measured one while it describes the data quite well for $\theta \gtrsim 50^\circ$.

The major systematic error in the calculated results comes from the uncertainty in the elementary cross section. Figures \ref{fig:Tpi} and \ref{fig:piang} also show in the dash-dotted curves the distributions calculated with the BNL input. As discussed earlier this is probably an overestimate of the true cross section and is shown here only in order to illustrate the effects of a change of input. Now the agreement between theory and experiment for the kinetic energy distribution (Fig.\ \ref{fig:Tpi}) is better (calculated cross section within all error bars). However, the angular distribution now becomes considerably worse (see Fig.\ \ref{fig:piang}) with a clear overshooting at the intermediate angles around $60^\circ$.

\subsubsection{Experiment-Theory discrepancies: Coherent Production} \label{sect:coh}
\begin{figure}[htb]
 \includegraphics[angle=-90,width=0.5\textwidth]{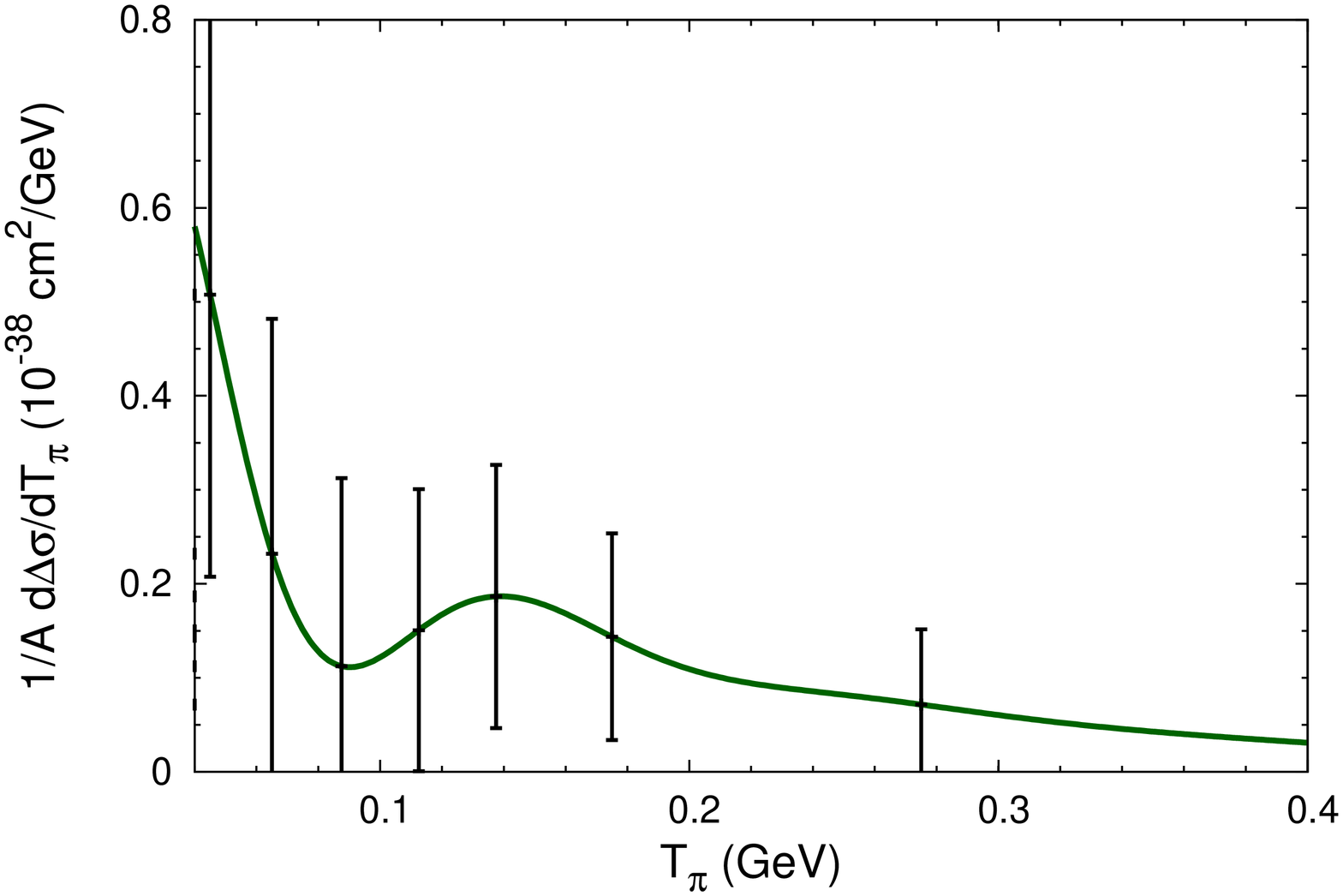} 
 \includegraphics[angle=-90,width=0.5\textwidth]{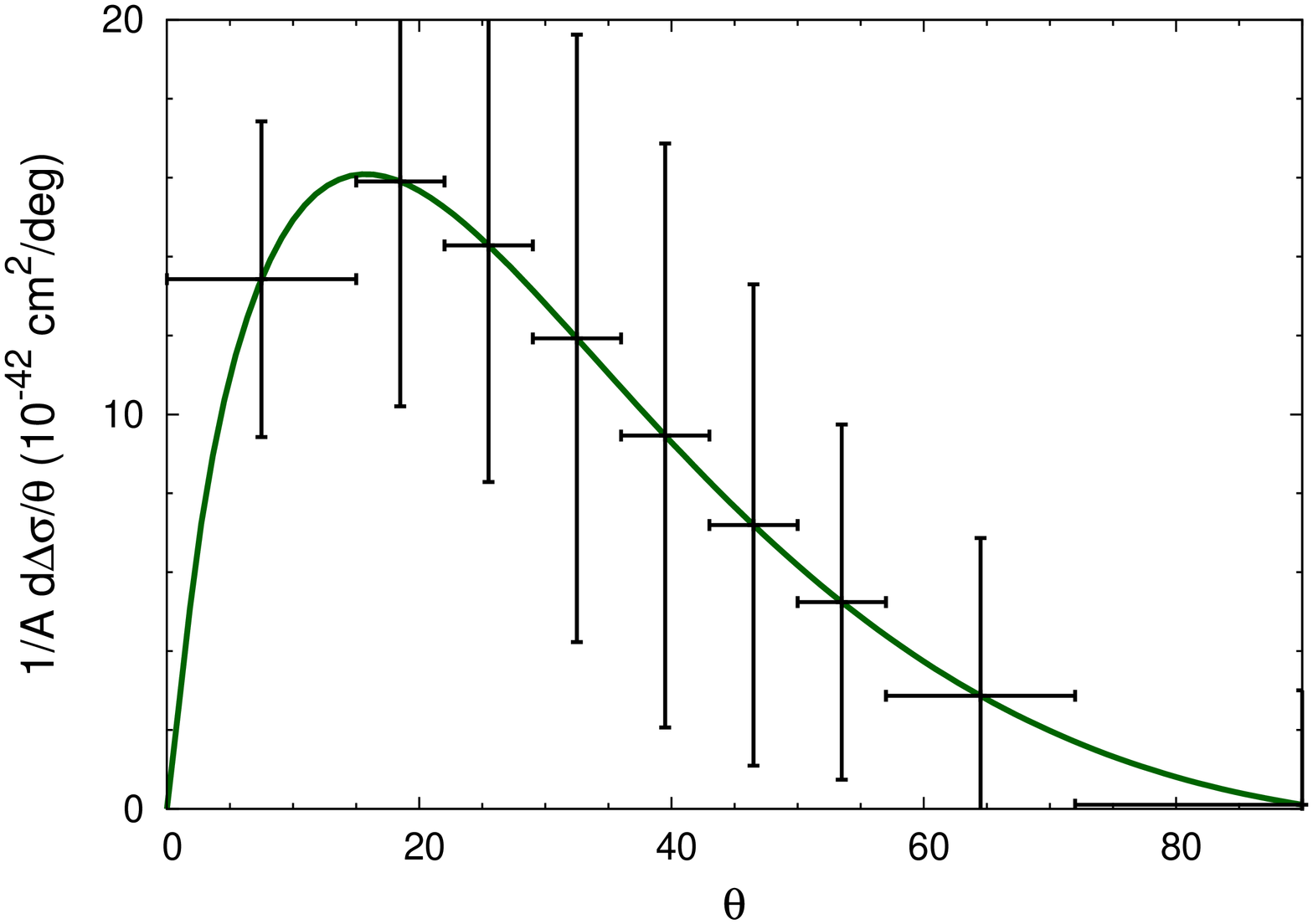}
 \caption{(color online)Difference of experimental and calculated kinetic energy (top) and angular distributions (bottom). The error bars are those of the data taken from \cite{Eberly:2014mra}. Note that in the upper figure the $T_\pi$ axis extends down to only 0.04 GeV. } \label{fig:picoh}
\end{figure}

In order to investigate the discrepancies between theory and experiment further the difference between a smoothed curve through the data for the kinetic energy distribution and the cross section calculated with the ANL input is shown in Fig.\ \ref{fig:picoh}. The error bars shown in that figure are only those given by Eberly et al.\ \cite{Eberly:2014mra} for the data; no systematic error for the GiBUU calculation has been added.
The difference in the kinetic energy distribution shows an interesting structure with a modest peak at about 0.14 GeV kinetic energy and a clear peak at the lowest energy of only about 0.04 GeV. The cross section amounts to roughly $0.15 \cdot 10^{-38}$ cm$^2$/GeV  for energies above about 0.1 GeV; at the energies below that the difference rises up to about $0.5 \cdot 10^{-38}$ cm$^2$/GeV at 0.04 GeV. Both the modest peak at 0.14 GeV and the sharp peak at 0.04 GeV lie above the error bars.

For the angular distribution (lower part of Fig.\ \ref{fig:picoh}) the difference between the calculated and the experimental values peaks at about $\theta = 18^\circ$. At the peak the error is about 30\%, while it increases to about 70\% for angles larger than about $\theta = 30^\circ$. Over all the angular range shown here up to about 60$^\circ$ the difference is significantly larger than 0.

GiBUU describes the incoherent pion production quite well. This assertion receives some support from the fact that it describes the angular distribution of the CC antineutrino-induced neutral pion production quite well as will be discussed in some more detail later in Sect.\ \ref{sect:antinupi0}. The excess at low kinetic energies and small angles, observed here for CC charged pion production, could then be due to the coherent production.

MINER$\nu$A has also measured the coherent cross section \cite{Higuera:2014azj}, but without the $W_{\rm rec} < 1.4$ GeV cut so that these data are not directly comparable with the ones extracted here. The measured value without the cut amounts to about $3.5 \times 10^{-39} \text{cm}^2$ \cite{Higuera:2014azj}; the integral over the difference shown in Fig.\ \ref{fig:picoh} amounts to about $1.9 \times 10^{-39} \text{cm}^2$ and is thus quite reasonable considering the different $W$-ranges. This is also supported by the experimental observation that the discrepancy at the lowest energy observed between the GENIE prediction and the experimental cross section amounts to about a factor of 1.5 and is nearly entirely due to events with $W > 1.4$ GeV \cite{MislivecFarland:2015}. A combined analysis both of the coherent  \cite{Higuera:2014azj} and the total \cite{Eberly:2014mra} data, using the same cuts and generator version, would help to clarify the situation.

Most generators use PCAC based models employing the Adler relations \cite{Morfin:2009zz}. The experimental analysis in \cite{Higuera:2014azj} finds 'poor agreement' of the models implemented in the standard generators GENIE and NEUT with the data; in particular the NEUT version gives a significantly broader (and correspondingly) larger coherent cross section. Recently, it has been argued that the Adler model breaks down for coherent excitations on nuclei \cite{Kopeliovich:2012kk} both at high and at low energies. In a detailed discussion of the Adler model it has also been shown that the original Rein-Sehgal (RS) as well as the Berger-Sehgal implementations of that model involve arbitrary assumptions \cite{Hernandez:2009vm}; in addition, the pion-nucleus elastic cross sections are not at all described by the RS model (see Fig.\ 2 in \cite{Hernandez:2009vm}), as implemented in GENIE \cite{Alvarez-Ruso:2015}. A complete theory of neutrino-induced coherent pion production must involve both reaction amplitudes connected with the peripheral $t$-channel scattering on the whole nucleus and, in addition, those connected with $s$-channel excitations of the nucleon resonances. This is well known from studies of coherent electroproduction of neutral pions at higher energies at JLAB \cite{Kaskulov:2011ab,Larin:2010kq} where the theory and the data exhibit both of these two contributions, one from the Primakoff effect and one from individual nucleon interactions.  The Adler model belongs to the first category; it suffers from the problems just discussed. Models for the resonance excitations are still under development \cite{AlvarezRuso:2007tt,Leitner:2009ph,Nakamura:2009iq}.

\subsubsection{Comparison with MiniBooNE CC pion production}
The MiniBooNE pion production data required input data even higher than the BNL values \cite{Lalakulich:2012cj} so that these data are obviously not compatible with the ones obtained by MINER$\nu$A. This disagreement between the MiniBooNE results and theory becomes larger with increasing for neutrino energies \cite{Lalakulich:2012cj}. Sobczyk and Zmuda \cite{Sobczyk:2014xza}, in a detailed comparison of both data sets, have recently pointed out that the experimental pion spectra of the MINER$\nu$A and the MiniBooNE experiment lie on top of each other for kinetic energies above about 125 MeV. At lower energy the MINER$\nu$A data are considerably higher so that the shapes of the kinetic energy distributions in both experiments are quite different.

\begin{figure}[htb]
 \includegraphics[angle=-90,width=0.5\textwidth]{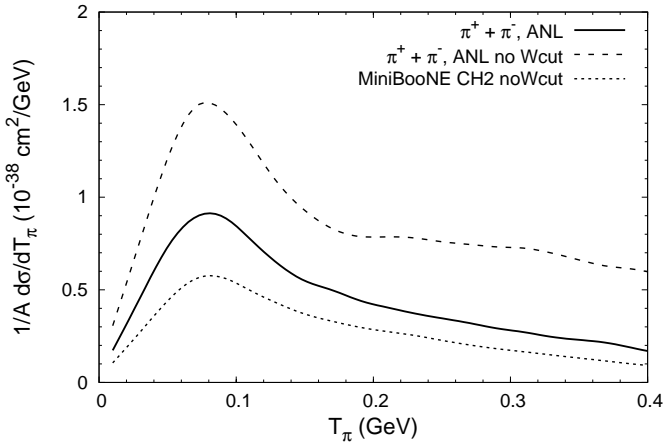}
 \caption{(Kinetic energy spectra for incoherent single charged pion production in CC reactions. The solid lines give the results for neutrino with the ANL input and a cut $W_{\rm rec} < 1.4$ GeV. The dashed, uppermost line shows the single-pi production spectrum without any $W_{\rm rec}$-cut. The dotted, lowest curve shows the results for the MiniBooNE flux and a CH$_2$ target, again without a $W_{\rm rec}$-cut.} \label{fig:picompMB}
\end{figure}

For comparison in Fig.\ \ref{fig:picompMB} also the calculated pion spectrum for the MiniBooNE experiment is shown as the lowest, dotted curve; for a more detailed presentation of GiBUU results for pion production in MiniBooNE see Ref.\ \cite{Lalakulich:2012cj}. This dotted curve should be compared to the middle solid line for MINER$\nu$A obtained with a $W$ cut at 1.4 GeV. The calculated MINER$\nu$A cross section is roughly by a factor 1.5 larger than that calculated with the MiniBooNE flux; this is in agreement with the comparison of results obtained with the NuWro generator in \cite{Sobczyk:2014xza}. However, the calculated shapes of the spectra  are quite similar, in contrast to the experimental behavior which shows the strong rise at low pion kinetic energies in the MiniBooNE experiment.

Neutrino-induced coherent pion production increases with energy and is localized at small kinetic energies. It is, therefore, tempting to assign the shape-difference to coherent excitation which is expected to be larger at MINER$\nu$A. Unfortunately, the theory of neutrino-induced coherent pion production is still in a preliminary stage (see discussion in the last section). The results shown in \cite{Higuera:2014azj} indicate an increase by about a factor of 2 between the lower MiniBooNE and the higher MINER$\nu$A energies.

\begin{figure}
\includegraphics[angle=-90,width=0.5\textwidth]{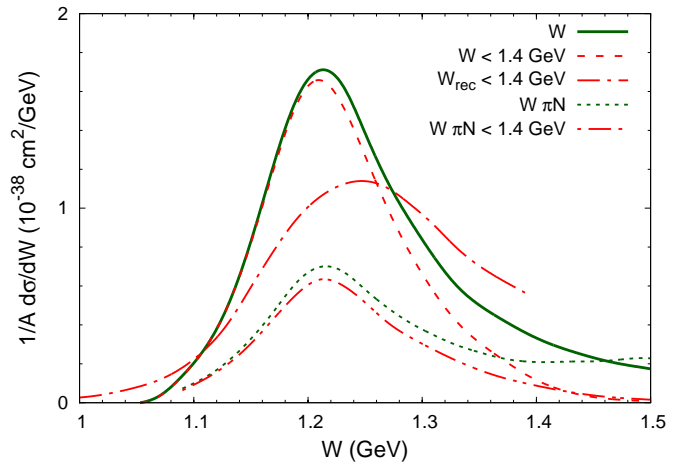}
\caption{(color online) Invariant mass distributions of incoherent events in which first a $\Delta$ resonance was excited. The calculations were performed for a $C$ target with the MINER$\nu$A flux. Shown are the true $W$-distribution without (solid, green curve) and with (dashed, red curve) cut. Also shown is the reconstructed $W_{\rm rec}$ distribution (red dash-dotted curve) and the distributions of the $\pi N$ invariant mass $W_{\pi N}$ without (dotted green curve) and with cut (red, dot-dot-dashed curve).} \label{fig:WDistr}
\end{figure}

\subsubsection{Invariant mass cuts}    \label{sect:W}

All the data in Figs.\ \ref{fig:Tpi} and \ref{fig:piang} were obtained with a cut on the reconstructed invariant-mass $W_{\rm rec} < 1.4$ GeV in the analysis \cite{Eberly:2014mra}. Here $W_{\rm rec}$  is defined as
\begin{equation} \label{Eq:Wrec}
W_{\rm rec}^2 = M^2 + 2M\omega - Q^2~,
\end{equation}
where $\omega$ is the energy transfer and $Q^2$ the four-momentum transfer. The experiment does not measure these latter two quantities directly but has to reconstruct them from a calorimetrically reconstructed incoming neutrino energy\footnote{It would be interesting to know how much of the final energy is actually measured and how much has to be reconstructed by using a generator.}. The cut has been used in \cite{Eberly:2014mra} to make the higher energy data obtained at MINER$\nu$A comparable to those obtained earlier at lower energies in the MiniBooNE experiment \cite{AguilarArevalo:2010bm} where the population of higher invariant masses was energetically suppressed.

In order to illustrate the effect of the cut on $W_{\rm rec}$ Fig.\ \ref{fig:picompMB} also shows the kinetic energy spectrum without any such cut (topmost dashed line). This uncut spectrum is larger by about a factor of 1.7 at the maximum. The cut on $W_{\rm rec}$ thus has a considerable influence on the published experimental cross section. This large effect is worrying since $W_{\rm rec}$ has no direct physical meaning for bound nucleons.

$W_{\rm rec}$ is the invariant mass for an interaction with a free, unbound nucleon at rest. For a bound and Fermi-moving nucleon the correct invariant mass is given by
\begin{equation}  \label{Eq:W}
W^2 = (E_N + \omega)^2 - (\mathbf{p_N} + \mathbf{q})^2  ~,
\end{equation}
where $E_N$ is the energy of the bound nucleon, $\mathbf{p_N}$ its momentum and $\mathbf{q}$ the three-momentum transfer. Also $W$
is experimentally not directly accessible.

Both $W_{\rm rec}$ and $W$ represent entrance channel properties. For pion production relevant are also the invariant mass distributions of the final $\pi N$ pairs
\begin{equation}     \label{Eq:WpiN}
W_{\pi N}^2 = (E_N + E_\pi)^2 - (\mathbf{p_N} + \mathbf{p_\pi})^2 ~.
\end{equation}
Here $E_\pi$ and $\mathbf{p_\pi}$ are the energy and the momentum of the pion. For values up to about 1.5 GeV $W_{\pi N}$ contains information on the $\Delta$ spectral function and is experimentally directly accessible.

In order to investigate these different definitions of an invariant mass in some more detail various $W$-distributions calculated with GiBUU are shown in Fig.\ \ref{fig:WDistr}. There is, first, the true $W$ distribution (Eq.\ \ref{Eq:W}), without any cut, shown by a solid green line.
 This curve has been obtained from a calculation without any cut. The same true $W$ distribution for a calculation in which the cut $W_{\rm rec} < 1.4$ GeV has been used is shown by the dashed red line. This curve shows no sharp cutoff at 1.4 GeV, but instead it looses strength on the high $W$ side, starting from the maximum\footnote{The peak in Fig.\ \ref{fig:WDistr} appears at a somewhat lower mass than the free $\Delta$ mass; this shift is due to the binding of the $\Delta$ inside the nucleus.} and upwards towards higher $W$. The cut on $W_{\rm rec}$, therefore, does not eliminate the pions from high-mass excitations, but instead distorts the spectral shape of the $\Delta$.

The distribution of $W_{\rm rec}$ (red, dashed-dotted curve) looks very different from the true $W$ distributions discussed so far. It is considerably lower in its peak cross section, broader and shifted to higher masses until it is cut off at 1.4 GeV. It shows no resemblance to the $\Delta$ spectral function.

Finally shown are also, by the green dotted and the red dot-dot-dashed lines, the invariant mass distributions of the final $\pi N$ pairs. Again, the red dot-dot-dashed curve obtained in a calculation with a cut on $W_{\rm rec}$ misses strength from 1.15 GeV on upwards. No sharp cut in $W_{\pi N}$ appears. Since the events in both distributions are subject to strong final state interactions they are lower than all the others.

\subsection{Charged pion production in the antineutrino beam}
\begin{figure}[htb]
 \includegraphics[angle=-90,width=0.5\textwidth]{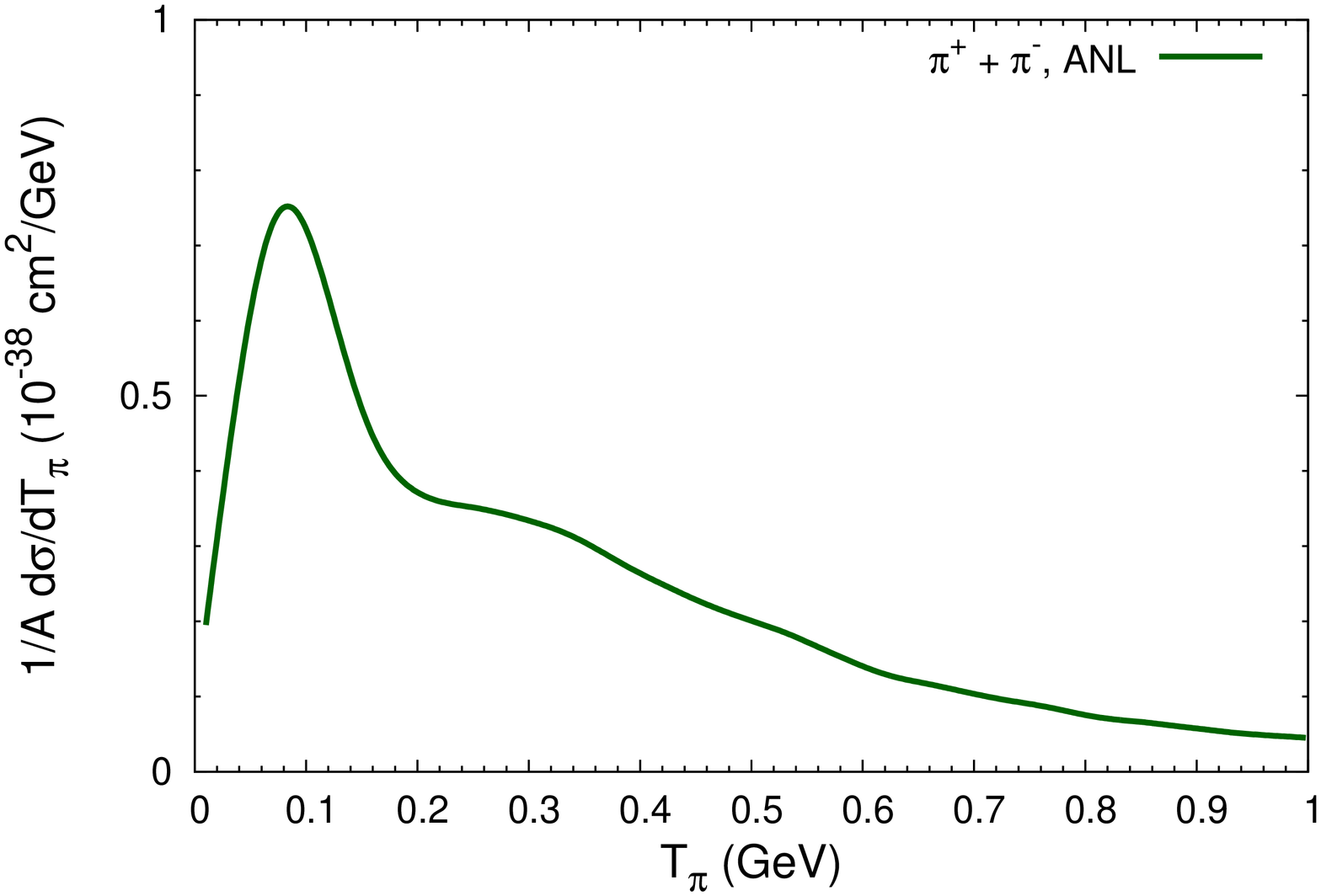}
 \caption{Kinetic energy spectra for incoherent single charged pion production in CC reactions with the MINERV$\nu$A antineutrino beam between 1.5 and 10 GeV. No $W_{\rm rec}$-cut has been used.} \label{fig:pi-antineutrino}
 \includegraphics[angle=-90,width=0.5\textwidth]{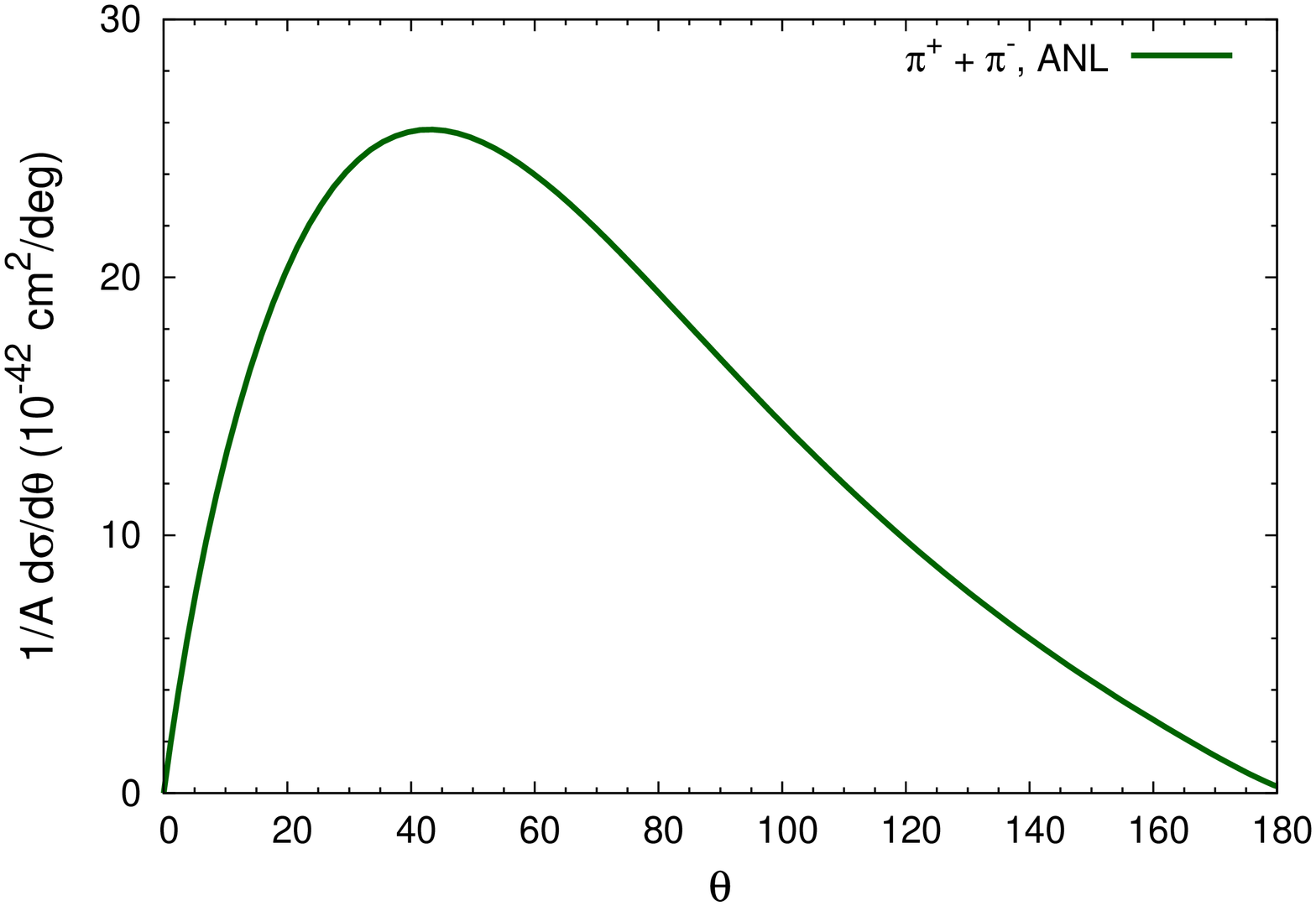}
 \caption{Angular distribution of incoherently produced charged pions in the MINER$\nu$A experiment with an antineutrino beam between 1.5 and 10 GeV. No $W_{\rm rec}$-cut has been used.} \label{fig:antinu_piang}
\end{figure}

For completeness in Fig.\ \ref{fig:pi-antineutrino} the charged pion spectrum (sum of $\pi^-$ and $\pi^+$) in an antineutrino beam is shown. These results were obtained with a flux $1.5 \:\:\text{GeV} < E_\nu < 10$ GeV and no invariant mass cut. The shape is very similar to that obtained for the neutrino beam (topmost curve in Fig.\ \ref{fig:picompMB}), but is lower by about a factor of 2 due to the different $V-A$ interference for antineutrinos.
In Fig.\ \ref{fig:antinu_piang} the angular distribution is given for the incoherent single pion production in the MINERV$\nu$A antineutrino beam.

\subsection{Neutral pion production in the neutrino and antineutrino beams}
The $\pi^0$ production both for neutrino and antineutrino induced CC reactions is interesting because there is no coherent excitation in this channel so that these results should be directly comparable to experiment. The GiBUU results discussed in this section also obtained with a flux 1.5 GeV$  < E_\nu < 10$ GeV and no invariant mass cut using the ANL input.

\begin{figure}[h]
 \includegraphics[angle=-90,width=0.5\textwidth]{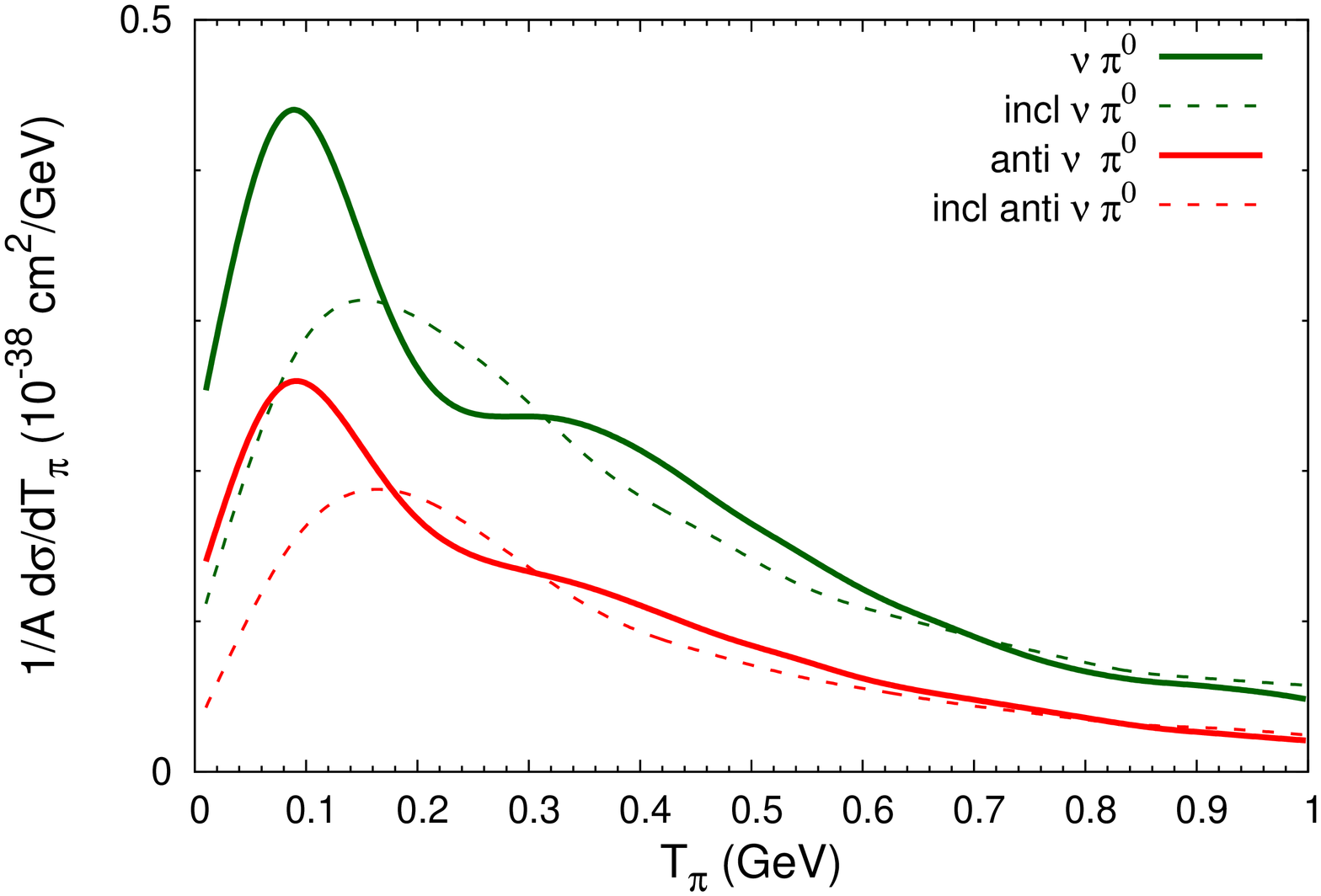}
 \caption{(color online) Kinetic energy spectra of incoherently produced single $\pi^0$ for CC reactions on a CH target using the MINERV$\nu$A neutrino and antineutrino fluxes between 1.5 and 10 GeV. The solid lines give the results for neutrino (upper, green) and for antineutrino (lower, red) beams with FSI included. The dashed lines, labeled with 'incl', give the corresponding results before FSI. No $W_{\rm rec}$ cut has been used.} \label{fig:pi0}
 \includegraphics[angle=-90,width=0.5\textwidth]{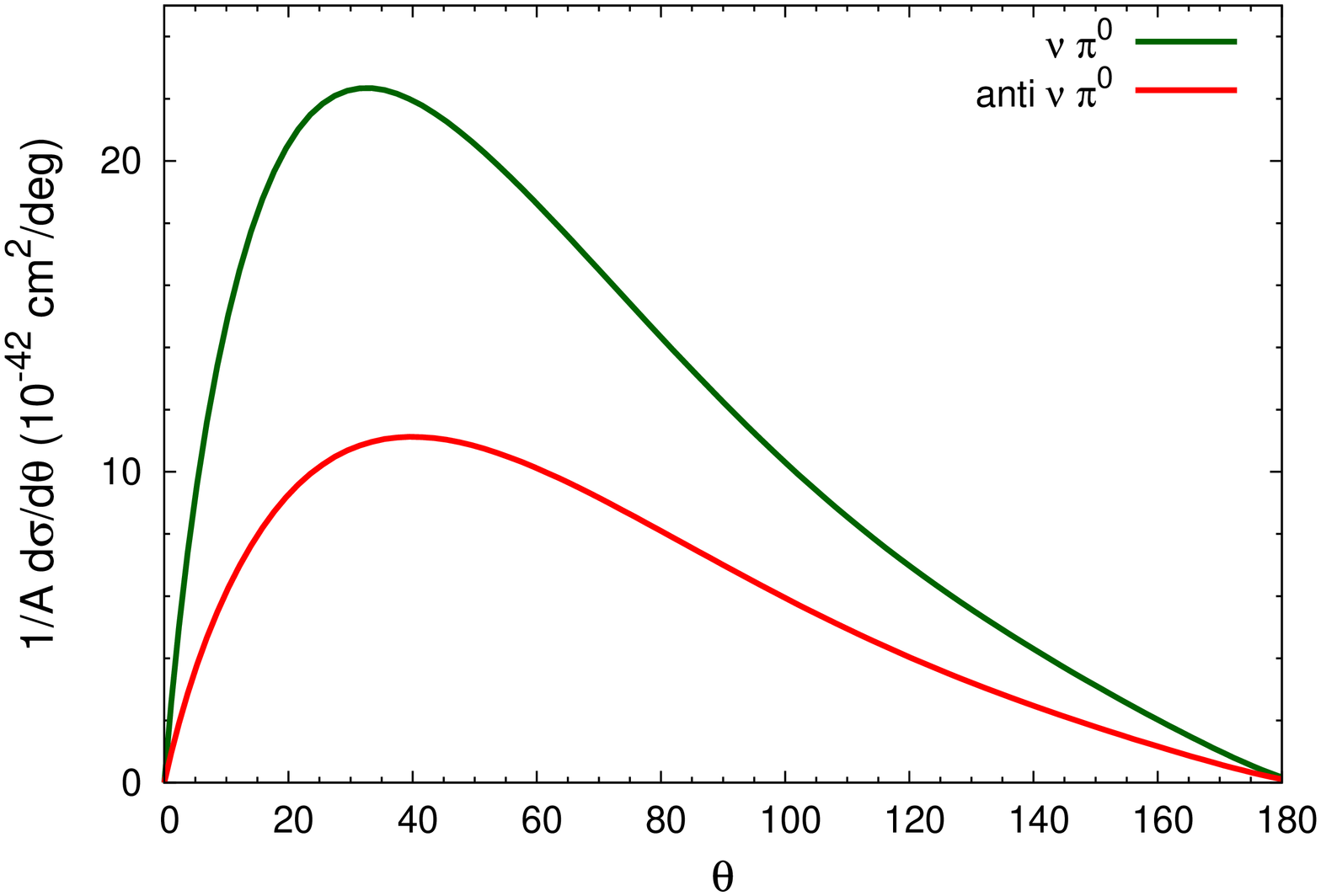}
 \caption{(color online) Angular distribution of incoherently produced single $\pi^0$ in the MINER$\nu$A experiment with the neutrino (green upper line) and antineutrino (red lower line) fluxes between 1.5 and 10 GeV. No $W_{\rm rec}$ cut has been used.} \label{fig:nu-antinu_piang}
\end{figure}

The two upper curves in Fig.\ \ref{fig:pi0} give the results for neutrino-induced and the two lower ones for antineutrino-induced neutral pion production. The effects of FSI are remarkable: the cross section after FSI are significantly larger than those before, in particular around 0.1 GeV kinetic energy. Furthermore the shape is significantly distorted by FSI.
The strong overshoot at 0.08 GeV is followed by an undershoot at around 0.24 GeV. For somewhat higher kinetic energies the cross section is again increased by the FSI and only from about 1 GeV on upwards there is a slight attenuation by final state interactions. The cross section after FSI shows a net increase compared to that before FSI. This increase is due to charge-transfer reactions from the dominant $\pi^+$ channel in the $\nu$-induced production and the $\pi^-$  channel in the $\bar{\nu}$ induced one. The importance of this charge transfer was already pointed out in \cite{Leitner:2006ww}; here it should be noted that GiBUU has been extensively tested against pion charge-exchange reactions on nuclei \cite{Buss:2006yk,Buss:2006vh}.

The region around $T_\pi = 0.24$ GeV reflects the strong pion absorption through the pion-less decay of the $\Delta$ resonance. This indicates that even at the high energies of the incoming neutrino beam in the GeV region, the $\Delta$ resonance plays an essential role in the FSI. Initially produced energetical pions cascade down through a sequence of elastic or inelastic scattering, possibly connected with charge transfer. The pions that finally end up in the $\Delta$ region can then be absorbed. Those pions that are slowed down even further can no longer be absorbed into a $\Delta$ thus causing the strong peak at 0.08 GeV.

The angular distributions for $\pi^0$ production shown in Fig.\ \ref{fig:nu-antinu_piang} look very similar to those for charged pion production.

\subsubsection{Neutral pion production at MINER$\nu$A}  \label{sect:antinupi0}
\begin{figure}[h]
 \includegraphics[angle=-90,width=0.5\textwidth]{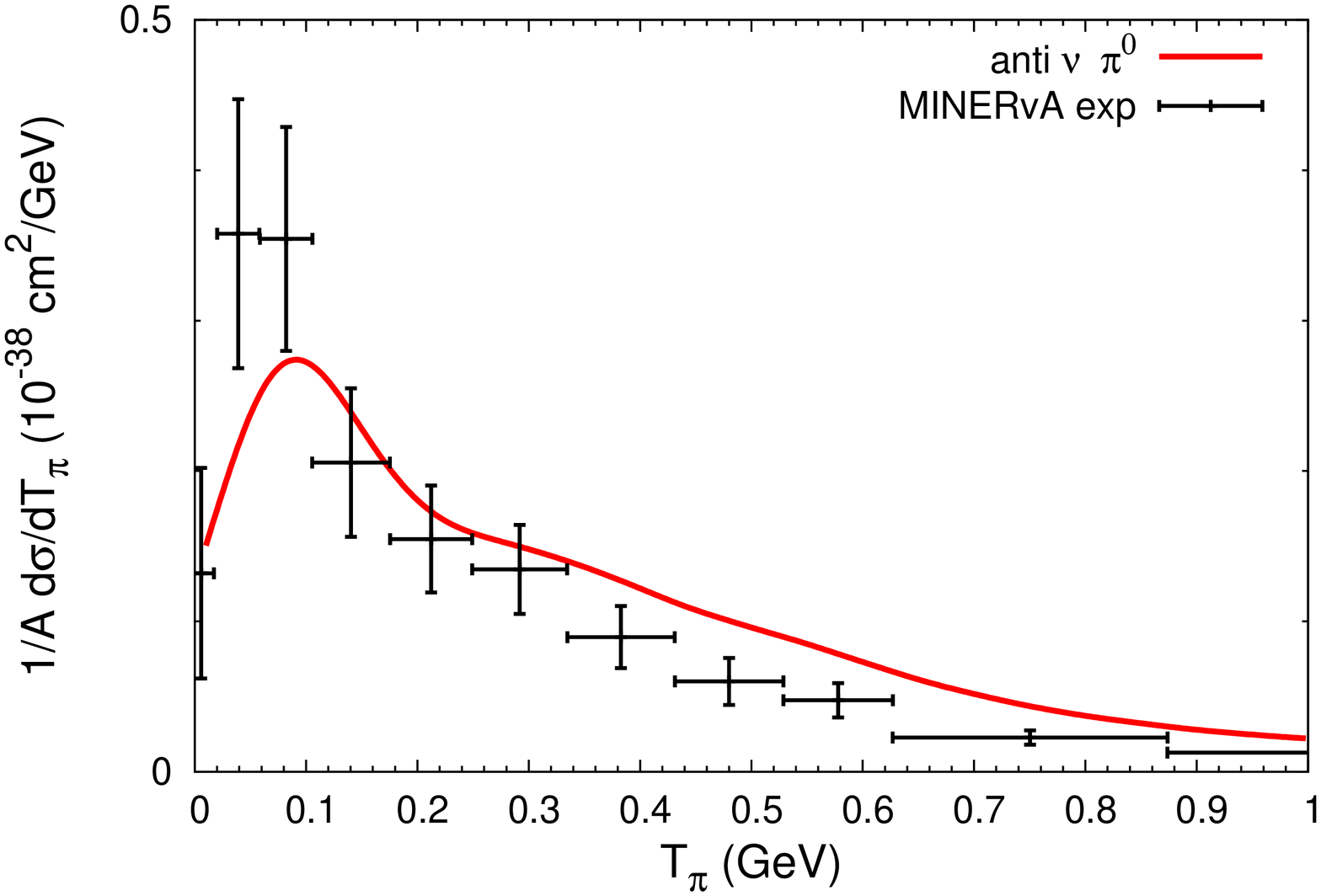}
 \caption{(color online) Kinetic energy spectra of incoherently produced single $\pi^0$ for CC reactions on a CH target using the MINERV$\nu$A antineutrino flux between 1.5 and 20 GeV. Data are from \cite{Aliaga:2015wva}, converted into $d\sigma/dT_\pi$.} \label{fig:antinupi0spectr}
 \includegraphics[angle=-90,width=0.5\textwidth]{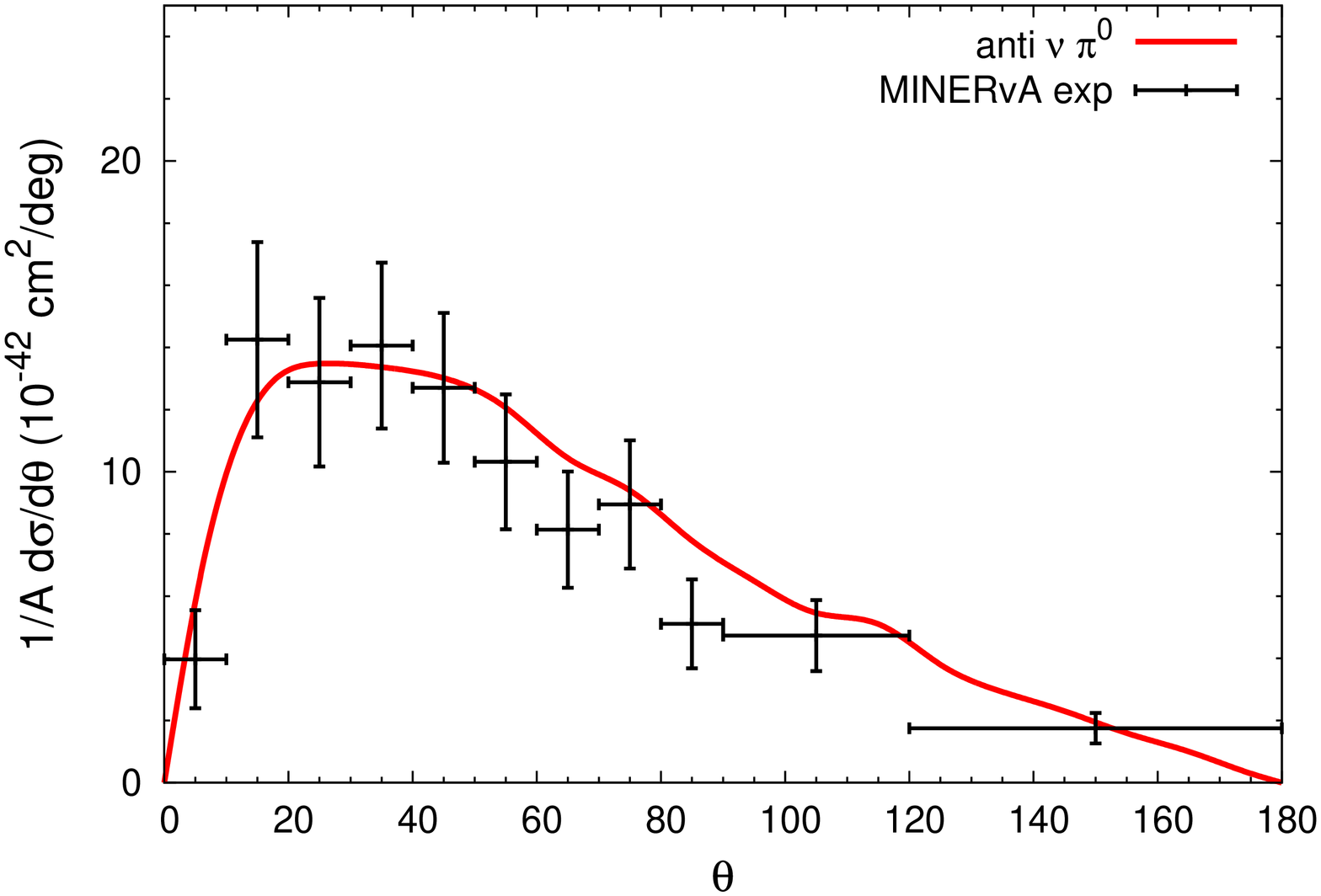}
 \caption{(color online) Angular distribution of incoherently produced single $\pi^0$ in the MINER$\nu$A experiment with the antineutrino flux between 1.5 and 20 GeV. Data are from \cite{Aliaga:2015wva}.} \label{fig:antinupi0ang}
\end{figure}

The calculated results for CC antineutrino-induced neutral pion production can be directly compared with data \cite{Aliaga:2015wva} that became available about 1 week after after a first version of the present paper paper had been uploaded to the arXiv \cite{Mosel:2015tja}. The main difference to the charged pion data, besides the different flavor of the incoming beam, consists in selecting the incoming neutrino energy to be between 1.5 and 20 GeV (instead of 10 GeV for the charged pions) and the absence of any $W$ cut. The absence of a $W$-cut has a more significant effect on the cross sections than the raising of the upper neutrino energy; the latter has about a 10\% effect. The results obtained with these these experimental specifications are shown in Figs. \ref{fig:antinupi0spectr} and \ref{fig:antinupi0ang}. Because of the absence of any CC coherent $\pi^0$ production of neutral pions these calculations are directly comparable with experiment.
The comparison of the experimental and theoretical angular distribution in Fig.\ \ref{fig:antinupi0ang} shows a very good agreement between both. In particular, there is no excess at small angles, as it was the case for the charged pion production discussed in earlier sections. This reflects the absence of any coherent production. The comparison of the calculated kinetic energy distribution with the data still exhibits an underprediction at the lowest kinetic energies around 40 and 80 MeV. On the other hand, for kinetic energies above about 0.3 GeV the data are lower than the theoretical prediction.

This behavior lends some support to the earlier discussions of a coherent contribution to the measured charged pion cross sections in Sect.\ \ref{sect:coh}.  The overshoot of the kinetic energy distribution at higher energies, combined with the underprediction at the lowest energies, may indicate an underestimate of FSI charge exchange reactions for the pion and/or problems in the experimental background subtraction.

\section{Summary and Conclusion}
In the energy regime of the planned DUNE pion production, either through resonances or DIS, represents the dominant reaction channel. It is, therefore, mandatory to obtain a quantitative understanding of this process. Quasi-elastic scattering, which has received a lot of theoretical attention recently, is necessarily entangled with pion production in the actual observables and can not be separated without the help of an event generator. Thus, the quality of the QE data can never be better than the quality of the generator used to remove the pion contamination. On the other hand, pion production initiated by neutrino interactions is interesting in itself since it may provide information on the axial couplings of resonances.

In this paper primarily the recent MINER$\nu$A data on charged pion production \cite{Eberly:2014mra} were analyzed. Unfortunately, the crucial input to such calculations, the pion production cross section on isolated nucleons, is still somewhat uncertain. While the discrepancy between two older data sets, from ANL and BNL, has recently found a reasonable explanation in terms of flux uncertainties in the BNL experiment \cite{Wilkinson:2014yfa}, there are still lingering problems with the extraction of nucleon cross sections from experiments using deuterium targets \cite{Wu:2014rga}. This latter problem clearly deserves more theoretical work, but ultimately a dedicated experiment using hydrogen targets is needed to clarify this point.

The calculations for charged pion production reproduce the measured kinetic energy spectra quite well, except for the lowest kinetic energies around 100 MeV, where the calculation comes out significantly lower than the experiment. This discrepancy can be localized in events connected with small scattering angles. At forward angles, the calculation significantly underestimates the experimental cross sections. These discrepancies, both in kinetic energy and angle, are compatible with contributions expected from a coherent excitation process. A reanalysis of the experimental MINER$\nu$A data on coherent pion production \cite{Higuera:2014azj}, using exactly the same cuts and generator tune as the Eberly et al analysis \cite{Eberly:2014mra}, would help to verify this explanation. This explanation also finds some support in the observation that the angular distribution of the charged current neutral pion production data in the antineutrino beam \cite{Aliaga:2015wva}, where no coherent component can be present, is reproduced quite well by the present calculations.

The comparison of the MiniBooNE pion production data with those obtained by MINER$\nu$A comes to the same conclusion as the recent analysis in \cite{Sobczyk:2014xza} that these two datasets are not compatible. The shape discrepancy may, however, reflect the presence of a coherent component in the spectra which will be larger at the higher energy of the MINER$\nu$A experiment. It is hoped that pion production data from the T2K experiment can shed some light on this problem \cite{Lalakulich:2013iaa}.

In an earlier paper \cite{Mosel:2014lja} the authors had already noted that the flux cut used in the MINER$\nu$A analyses introduces a model dependence into the comparison of theory with data. The cuts used in theoretical calculations are those for true energies whereas the experimental cuts can only be done for reconstructed energies. It was shown in  \cite{Mosel:2014lja} that the flux cuts used by the MINER$\nu$A experiment can significantly distort the event rates, requiring a fairly large and generator-dependent correction through energy-migration matrices.  Here now the effects of cuts imposed also on the reconstructed invariant mass $W_{\rm rec}$ in obtaining the pion production data were investigated. The cut has a major influence on the total pion production cross section. This was to be expected since a large part of the pions is produced by DIS events. A cut at 1.4 GeV, as employed in the experiment, was intended to remove these contributions and enrich the $\Delta$ resonance contribution. This, however, is not what the $W_{\rm rec}$ cut actually does. Instead it cuts off strength over a large, high-mass part of the $\Delta$ spectral function, starting already around its peak value. The explanation for this behavior lies in the Fermi-motion and binding energy of the nucleons inside the nuclear target. A comparison of pion cross sections obtained with this cutoff on $W_{\rm rec}$ with data in other experiments and at lower energies is then difficult. In addition, this cut -- being an entrance channel cut -- has to rely on the reconstruction of neutrino energy which introduces a model dependence into the data. A cut on the $\pi -N$ invariant mass in the final state would be free of this problem and is thus preferable.

\begin{acknowledgments}
I am grateful to Kevin McFarland and to Michael Wilking for many helpful discussion on neutrino interactions and pion production in particular. Discussions with Luis Alvarez-Ruso on coherent pion production were very useful.

 I also gratefully acknowledge the help and support of the whole GiBUU team in developing both the physics and the code used here. Here, in particular Kai Gallmeister and Janus Weil have been most helpful with computational problems. My thanks further go to Tina Leitner and Olga Lalakulich who developed large parts of the code used for this study.

This work has been supported by Deutsche Forschungsgemeinschaft (DFG).
\end{acknowledgments}

\bibliographystyle{apsrev4-1}
\bibliography{nuclear}

\end{document}